\documentclass{article}
\usepackage{amsmath}
\begin{document}

\date{}
\title{NON--INTEGRABILITY OF A WEAKLY INTEGRABLE HAMILTONIAN SYSTEM}

\author{Giuseppe Pucacco\thanks{e-mail: pucacco@roma2.infn.it} \\ 
Dipartimento di Fisica -- Universit\`a di Roma ``Tor Vergata" \\ 
and\\
INFN -- Sezione di Roma II\\
Kjell Rosquist\thanks{e-mail: kr@physto.se} \\
Department of Physics -- Stockholm University}

\maketitle

\begin{abstract}
The geometric approach to mechanics based on the Jacobi metric allows                         
to easily construct  natural mechanical systems which are integrable (actually 
separable) at a  fixed value of the energy. The aim of the present paper 
is to investigate the dynamics of a simple prototype system
outside the zero-energy hypersurface.  We find that the
general situation is that in  which integrability is not
preserved at arbitrary values of the energy.  The structure
of the Hamiltonian in the separating coordinates at zero
energy  allows a perturbation treatment of this system at
energies slightly different  from zero, by which we obtain
an analytical proof of non-integrability.
\end{abstract}

\clearpage

%%%%%%%%%%%%%%%%%%%%%%%%%%%%%%%%%%%%%%%%%%%%%%%%%%%%%%%%%%%%%%%%%%%%%%%%%%%%%%

\section{Introduction}

The property of a Hamiltonian system to be integrable may happen to be
satisfied only at a fixed value of the energy. In this case we can speak 
of {\it
weak integrability} and refer to the phase-space functions which are 
conserved only in
correspondence of those given energy values as {\it weak invariants}.  
From this point of view,
standard integrability, with invariants which are conserved functions at
arbitrary energies, can be referred to as {\it strong} integrability.

Weak invariants, also called {\em configurational invariants}, have been
discussed by Hall (1983) and by Sarlet, Leach and Cantrijn
(1985).  Hietarinta (1987), in his account of the
direct methods for the search of the second invariant, also provides a 
review
of all the known 2-dimensional systems admitting one or more 
configurational
invariants.  In the present paper we use the approach of Rosquist and 
Pucacco 
(1995), where quadratic invariants at arbitrary and fixed energy for
2-dimensional systems were treated in a unified way. As shown in that 
paper,
the integrability condition for quadratic
invariants, corresponding to second-rank Killing tensors of the conformal
Riemannian geometry with Jacobi metric, involves an arbitrary analytic
function $S(z)$. For invariants at arbitrary energy, the function
$S(z)$ is a second degree polynomial with real second derivative and the
integrability condition then reduces to the classical Darboux's condition 
for
quadratic invariants at arbitrary energy (Darboux, 1901; Whittaker, 1937). 
The
possibility of searching for linear and quadratic invariants at fixed 
energy was
also addressed and some examples of systems admitting a second quadratic
invariant at zero energy were provided.

Generalizing the approach, Karlovini and Rosquist (2000) have discussed
the existence of invariants {\em cubic in the momenta} at both fixed and
arbitrary energy.  Besides giving a list of all known systems admitting a 
cubic
strong invariant, they find a superintegrable system admitting a cubic
configurational invariant related to an energy dependent linear 
invariant.  In
Karlovini, Pucacco, Rosquist and Samuelson (2002) we discuss the case of 
{\em quartic}
invariants associated with the existence of fourth-rank Killing tensors. 
The
results obtained in this geometric framework are in agreement with the 
direct
method when the class of potentials examined is the same, as can be seen
comparing the cases listed in the above papers with those appearing in 
Nakagawa
and Yoshida (2001).

The set of weakly integrable systems is very large but, on the other hand,
they are probably of limited usefulness in physical applications.  One may
wonder however if the knowledge of a weak invariant can provide information
about the global dynamical behaviour of the system.  One possibility is to
investigate the existence of integrable systems with a higher-order strong
invariant related to the weak one (e.g. linear in the momenta, as in one 
example
given in Karlovini and Rosquist, 2000). The aim of the present paper is 
instead to
explore the phase-space structure of the system at energies different from
that assuring weak integrability in the general situation in which the 
system is
non-integrable. In one class of systems, we show how the property of weak
integrability leads to a better understanding of the dynamics of generic
systems. In particular, the setting in which the geometric approach casts 
the
problem allows to apply powerful tools of analytical mechanics to prove the
non-integrability of the dynamical system.

The plan of the paper is as follows: in section 2 we recall conditions for 
weak 
and strong integrability of a two-dimensional natural Hamiltonian system
limiting the application to the existence of a second invariant quadratic 
in the
momenta; in section 3 we present a class of weakly integrable systems 
which have
a quasi-regular behavior at near-zero energy and for which we give a Poincar\'e 
type argument
of non integrability out of the zero energy surface; section 4 contains the conclusions.

%%%%%%%%%%%%%%%%%%%%%%%%%%%%%%%%%%%%%%%%%%%%%%%%%%%%%%%%%%%%%%%%%%%%%%%%%%%%%%

\section{Integrability at fixed energy}

The technique devised and applied in Rosquist and Pucacco 
(1995, RP here and in the following), allows to find the conditions such 
that a
2-dimensional natural Hamiltonian system admits a second invariant 
quadratic in
the momenta. In this section we briefly recall these results that are at 
the basis
of the applications of the following sections.

%%%%%%%%%%%%%%%%%%%%%%%%%%%%%%%%%%%%%%%%%%%%%%%%%%%%%%%%%%%%%%%%%%%%%%%%%%%%%%

\subsection{Weak quadratic invariants}

We are interested in the classical 2-dimensional systems with Hamiltonian
function
\begin{equation}\label{H}
{\cal H} = \tfrac12 (p_x^2 + p_y^2) + V (x, y).
\end{equation}
Since the Hamiltonian is time-independent, energy is conserved and motion
takes place on the hypersurface
\begin{equation}\label{Hyper}
{\cal H} (p_x, p_y, x, y) = E.
\end{equation}
The approach of RP amounts to find a conformal transformation to new 
coordinates $X,Y$ defined
by
\begin{equation}\label{CC}
z = F(w) , \quad z = x + i y, \quad w =  X + i Y , 
\end{equation}
generated by an arbitrary analytic function $S(z)$ via the relation
\begin{equation}\label{dct}
F' (w(z)) = \sqrt{S(z)}.
\end{equation}
The transformation (\ref{CC}) naturally induces a canonical point 
transformation that
gives the new Hamiltonian
\begin{equation}\label{SH}
{{\cal H}}_S = \frac{\tfrac12 (p_X^2 + p_Y^2) }{|S(X,Y)|} + V(X,Y) \ .
\end{equation}
where it appears the conformal factor 
\begin{equation}\label{confact}
|S(X,Y)| = \sqrt{ S(w) \bar S (\bar w)} = F' (w) \bar F' (\bar w).
\end{equation}
One can then show that, if the
function
\begin{equation}\label{G}
G = E - V,
\end{equation}
the so-called ``Jacobi" potential, can be expressed, in the new 
coordinates $X,Y$, in
the form
\begin{equation}\label{GAB}
G(X,Y) = \frac{A(X;E) + B(Y;E)}{|S(X,Y)|} ,
\end{equation}
where $A$ and $B$ are arbitrary functions of their arguments, the
quadratic function
\begin{equation}\label{JSI}
I_0 (p_X, p_Y, X, Y) = \tfrac12 (p_X^2 - p_Y^2) - A(X) + B(Y) ,
\end{equation}
commutes with the Hamiltonian (\ref{SH}) on the surface 
\begin{equation}\label{zero}
{\cal H} (p_x, p_y, x, y) = 0 .
\end{equation}
We therefore speak of a {\it
weak} invariant of the standard Hamiltonian system, in contraposition with 
the
usual notion of {\it strong} invariant by which we mean a phase-space
function which is conserved at arbitrary values of the energy. 

The origin of the phenomenon of weak integrability in the present context 
is
due to the geometrization of the dynamics via the Jacobi-Maupertuis
variational principle.  For a detailed account of the Jacobi 
geometrization we
refer to standard textbooks (Abraham and Marsden, 1978; Arnold, 1978; 
Lanczos,
1986) and to RP where, in particular, the link is
deepened between quadratic invariants and second-rank Killing tensors 
which is at
the basis of this approach. Here we briefly recall that to the Jacobi 
potential (\ref{G}) pertains a
family of conformal Riemannian metrics parametrized by the value of the 
energy. 
Each energy specifies a geodesic flow on a Riemannian manifold, whose 
projections
coincide with the orbits of the system defined by the standard Hamiltonian 
up
to a time reparametrization.  The solution of the Killing tensor equations
provides an invariant along the given flow at a given value of the energy
parameter (in particular at {\it zero} energy).  It is only the additional 
requirement 
that the solution of Killing equations be independent of the
energy that leads to integrability at arbitrary energy and poses a further 
constraint on the generating function $S(z)$. If this condition is 
satisfied, the function (\ref{JSI}) commutes with the Hamiltonian 
(\ref{SH}) on the surface (\ref{Hyper}), $\forall E$. One can clearly 
also extend the investigation of higher rank Killing tensors or more 
general functional forms of the second invariant and to the corresponding 
analysis of the conditions for strong integrability.

%%%%%%%%%%%%%%%%%%%%%%%%%%%%%%%%%%%%%%%%%%%%%%%%%%%%%%%%%%%%%%%%%%%%%%%%%%%%%%

\subsection{Strong quadratic invariants}

In RP it has been shown that, to get quadratic invariants at arbitrary 
energy, the
function $S(z)$ must satisfy the condition
\begin{equation}\label{S2}
{\rm Im} \{ S''(z) \} = 0 ,
\end{equation}
that is it must be a second degree polynomial with real second derivative.
Actually, we can show that condition (\ref{S2}) not only implies strong 
integrability, but also assures the separation of variables. We have in 
fact the following

\vskip.3cm

\noindent
{\bf Theorem 1:} Given the conformal transformation
\begin{equation}
d z = F'(w) \ d w = \sqrt{ S(z(w))} dw,
\end{equation}
condition necessary and sufficient to have 
\begin{equation}\label{separata}
|S(X,Y)| = A_S (X) + B_S (Y), 
\end{equation}
where $A_S$ and $B_S$ are real analytic functions of a single variable, is 
that (\ref{S2}) is true.

\vskip.3cm

{\it Proof:} Let us prove that (\ref{separata}) implies (\ref{S2}).
Functional form (\ref{separata}) is equivalent to say that the function 
$|S|$ must satisfy the
differential equation
\begin{equation}\label{diffeq}
|S|_{,XY} = 0.
\end{equation}
But
\begin{eqnarray}
|S|_{,XY} &=& i 
\bigl[ |S|_{,ww} - |\bar S|_{, \bar w \bar w} \bigr] = i 
\bigl[ \sqrt{S} \bigl(\sqrt{S} (\sqrt{S \bar S})_{,z} \bigr)_{,z} -  
 \sqrt{\bar S} \bigl(\sqrt{\bar S} 
(\sqrt{S \bar S})_{,\bar z} \bigr)_{,\bar z} \bigr]\\
 &=& i |S| \bigl( S_{,zz} - \bar S_{, \bar z \bar z} \bigr) = 
- 2 |S| \ {\rm Im} \{ S'' \}.
\end{eqnarray}
Since $|S|$ is a non-null positive function everywhere, the implication is 
evident. On the
contrary, (\ref{S2}) is equal to
\begin{equation}\label{SS2}
S = a z^2 + \beta z + \gamma,
\end{equation}
with $a$ real and $\beta, \gamma$ complex. This implies (\ref{separata}) 
as can be verified by
performing the coordinate transformation \eqref{CC}, with
\begin{equation}\label{FS2}
w = F^{-1}(z) = \int \frac{dz}{\sqrt{a z^2 + \beta z + \gamma}},
\end{equation}
where \eqref{SS2} has been used in defining the generating function via 
\eqref{dct}. Q.E.D.

\vskip.3cm

In RP, Sect. 4, the explicit forms of functions $A_S$ and $B_S$ are given. 
They can be of the
four possible types corresponding to separability in Cartesian, polar, 
parabolic and elliptic
coordinates. In this way, if we write the two arbitrary functions 
appearing in (\ref{GAB}) as
\begin{eqnarray}
A(X;E) &=& E A_S (X) - f(X),\\
         B(Y;E) &=& E B_S (Y) - g(Y),
\end{eqnarray}
the ``true" potential takes the form
\begin{equation}\label{VFG}
V(X,Y) = \frac{f(X) + g(Y)}{|S(X,Y)|} ,
\end{equation}
and the
Hamilton-Jacobi equation for Hamilton's characteristic function, $ {\cal 
W} (X,Y)
$, associated to Hamiltonian (\ref{SH}), takes on the explicitly separated 
form
\begin{equation}
\tfrac12 \bigl[({{\cal W}_{,X}})^2 + ({{\cal W}_{,Y}})^2 \bigr]
      + f(X) + g(Y) - E \bigl[ A_S (X) + B_S (Y) \bigr] = 0 ,
\end{equation}
and therefore strongly integrable two-dimensional systems with quadratic
second invariants can be exhaustively classified. 

%%%%%%%%%%%%%%%%%%%%%%%%%%%%%%%%%%%%%%%%%%%%%%%%%%%%%%%%%%%%%%%%%%%%%%%%%%%%%%

\subsection{Regularization}

Before starting the analysis of weakly integrable system, it is useful to
recall the fact that the conformal transformation leads, in a natural way, 
to
a time reparametrization of the dynamics which can be usefully exploited in
the applications.  This reparametrization has been used since a long time 
in
the framework of the ``regularization" of singular differential equations,
especially in the applications in celestial mechanics (see, e.g.
Levi-C\`{\i}vita, 1956; Sundman, 1912).  We can therefore speak of a 
generalized
regularization approach and the null Hamiltonian introduced below can be 
called on
the same footing the {\it regularized} Hamiltonian.

Let us suppose that we have chosen a function $S(z)$ which generates the
conformal transformation which gives a system admitting a weak second
invariant (\ref{JSI}). If we denote by
${{\cal H}}_0$ the function
\begin{equation}\label{Hzero}
\tfrac12 (p_X^2 + p_Y^2) + f(X) + g(Y) ,
\end{equation}
which can be interpreted as the numerator appearing in the Hamiltonian 
(\ref{SH}) once the expression (\ref{VFG}) is taken into account, the 
Poisson brackets of a generic phase-space function $J$ with
${{\cal H}}_S$ can be written as
\begin{equation}\label{poisson}
      \{J,{\cal H}_S \} = \frac{1}{|S|} \{J, {{\cal H}}_0 \}
      - \frac{{\cal H}_0}{|S|^2} \{J,|S| \}
      =\frac{{\{J, {\cal H}_0 \} - E \{J, |S| \}} }{ {|S|}}. \
\end{equation}
On the other hand, let us introduce the
{\it null} Hamiltonian
\begin{equation}\label{nullHam}
{{\cal H}}_{\cal N} = ({{\cal H}}_S - E) |S| =
                     {{\cal H}}_0 - E  |S| .
\end{equation}
The Poisson bracket of a function with ${{\cal H}}_{\cal N}$ is given by
\begin{equation}\label{poisson2}
\{J, {{\cal H}}_{\cal N} \} =
\{J, {{\cal H}}_0 \} - E \{J, |S| \} .
\end{equation}
Eqs.(\ref{poisson})
and (\ref{poisson2}) are therefore equivalent with respect to conservation 
of
$J$.  The conceptual difference relates to the fact that, whereas 
vanishing of
eq.(\ref{poisson2}) expresses the possible conservation of $J$ in the 
dynamics provided
by ${{\cal H}}_{\cal N}$ at {\it zero ``energy"} (i.e.\ ${\cal H}_{\cal 
N}=0$), the
vanishing of eq.(\ref{poisson}) expresses the conservation of $J$ in the
dynamics provided by ${{\cal H}}_S$ at {\it arbitrary} energy.  Note that 
the
physical energy $E$ enters into ${\cal H}_{\cal N}$ as an arbitrary 
parameter.

With the null Hamiltonian, it is automatically introduced a new 
(regularizing)
time variable $\eta$ by means of
\begin{equation}\label{newtime}
d \eta = \frac{d t }{ |S|} .
\end{equation}
Just as the Poisson brackets (\ref{poisson}) express the total derivative
with respect to the standard time
\begin{equation}\label{oldder}
\frac{d J }{d t} = \{J, {{\cal H}}_S \} ,
\end{equation}
the Poisson brackets (\ref{poisson2}) express the total derivative with
respect to the regularizing time
\begin{equation}\label{newder}
\frac{d J}{d\eta} = \{J, {{\cal H}}_{\cal N} \} .
\end{equation}
Correspondingly, the ``regularized" equations of motion are
\begin{eqnarray}\label{motions}
\frac{d X}{d\eta}  &=& \{X, {{\cal H}}_{\cal N}   \} , \qquad
         \frac{d p_X}{d\eta} = \{p_X, {{\cal H}}_{\cal N} \} , \\
         \frac{d Y}{d\eta}  &=& \{Y, {{\cal H}}_{\cal N}   \} , \qquad
         \frac{d p_Y}{d\eta} = \{p_Y, {{\cal H}}_{\cal N} \} . 
\end{eqnarray}

It is worthwhile to remark that ${\cal H}_S$ gives the same dynamics as the original
physical Hamiltonian ${\cal H}$. Only different coordinates are used as a result of a 
canonical point transformation. The time reparametrization implies the relation between
Poisson brackets associated with the original physical Hamiltonian and the null regularized
one. Considering the phase-space function of the regularized system as $J(p_X,
p_Y, X, Y; E)$ and the corresponding phase-space function of the original system, say 
$I(p_x, p_y, x, y)$, obtained via the recipe
$$
\{p_X,p_Y,X,Y\} \rightarrow \{p_x,p_y,x,y\}, \quad
E \rightarrow {\cal H} (p_x,p_y,x,y),
$$
the total time-derivative along the flow of these two functions are related by
$$
\frac{d J}{d\eta} = |S| \frac{d I}{d t} .
$$
The conformal factor is never vanishing and therefore we extend considerations above on
conserved quantities to functions in the original coordinate frame.

It should also be recalled that the combination of the conformal transformation
(\ref{CC}) with the introduction of the new time variable by means of
(\ref{newtime}) can also be seen as a canonical transformation on the 
extended
phase space where ${\cal H}$ and $t$ are new canonical coordinates
(Tsiganov, 2000).

%%%%%%%%%%%%%%%%%%%%%%%%%%%%%%%%%%%%%%%%%%%%%%%%%%%%%%%%%%%%%%%%%%%%%%%%%%%%%%

\subsection{Dynamics at zero energy}

The dynamics provided by the null or regularized Hamiltonian 
(\ref{nullHam})
is particularly simple in the separable case: the two motions in the
separating coordinates $X$ and $Y$ decouple and the general motion is 
given by
a superposition with independent arbitrary initial conditions.  Clearly, 
the
same happens in the even simpler case of a system constructed as above
choosing an arbitrary conformal transformation and selecting only motions 
at
zero energy.

In the $\eta$ time, the equations of motion (\ref{motions}) given by 
Hamiltonian (\ref{Hzero}) are simply
\begin{eqnarray}
\frac{d^2 X}{d\eta^2} &=& -f'(X) , \\
         \frac{d^2 Y}{d\eta^2} &=& -g'(Y) . 
\end{eqnarray}
The two ``energy" equations
\begin{eqnarray}\label{energies}
\tfrac12 p_X^2 + f(X) &=&h_1 , \\
         \tfrac12 p_Y^2 + g(Y) &=&h_2 , 
\end{eqnarray}
with
\begin{equation}
h_1 + h_2 = 0 ,
\end{equation}
delimit the regions admitted to the motion by means of the inequalities
\begin{eqnarray}\label{regions}
f(X) & \le & h_1 , \\
         g(Y) & \le & h_2 . 
\end{eqnarray}
From the theory of conformal           
transformations (see, e.g., Markushevitch, 
1983),
the ``isothermal" net of resulting  
coordinates is orthogonal (except in 
the finite
number of isolated singularities where 
$F'=0$).  Since in the general 
case, the
frequencies of the two motions in (\ref{motions}) are incommensurable, the 
regions
admitted by the motion, with coordinates satisfying inequalities 
(\ref{regions}),
are densely filled by the representative point.  This happens both in the 
presence
of rotations and librations with the two possibilities determined by the 
explicit
form of the potential functions $A$ and $B$ and by initial conditions.  
This
picture is the same even if we describe the dynamics with the original 
Hamiltonian
${{\cal H}}_S$.  What changes is only the speed of the representative 
point along the
trajectory.

Due to the almost trivial nature of the dynamics, we would like to know if 
we
can exploit this simplicity to gain information about the much more complex
situation of arbitrary energy.  In the hypothesis of integrability at zero
energy, at energies different from zero there are two possibilities: either
the integrability of the system is preserved or it is broken.  In the first
case, a second strong invariant exists but there is no general procedure to
find it.  Isolated systems with second invariants which are polynomial in 
the
momenta can be identified with the techniques described in the papers 
cited in the
Introduction. However, the general case we expect is that in which 
integrability
breaks down.  This is the subject of the rest of the paper, where we 
address the
following question: can we use the setting developed so far to investigate 
and
possibly predict the non-integrability of a given system?

Before closing this section, we would however mention the fact that, due to
the possibility of quite involved conformal transformations, a motion that
looks trivial in the separating variables, may become complicated with many
peculiar features when displayed in the physical coordinates of the 
original
problem.  Therefore, even the case with integrability limited to the
hypersurface of zero energy can have direct practical applications.

%%%%%%%%%%%%%%%%%%%%%%%%%%%%%%%%%%%%%%%%%%%%%%%%%%%%%%%%%%%%%%%%%%%%%%%%%%%%%%

\section{Systems of the class $S(z) = i z^2$ }

We start now the analysis of 2-dimensional systems generated via general
conformal transformations that do not stay in the restricted class 
specified
by Theorem 1.  These systems are therefore integrable (actually
separable) on the zero-energy surface.  We have selected a set of systems 
for
which the question posed in the previous section can be given a meaningful
answer, with the criteria of the simplest non-trivial situations.

We consider the function
\begin{equation}\label{iz2}
S(z) = i z^2 .
\end{equation}
This is the simplest polynomial which does not satisfy the constraint
(\ref{S2}) and therefore gives a potential which is not automatically
integrable at arbitrary energy.

Let us examine the coordinate transformation generated by (\ref{iz2}).  By
virtue of (\ref{dct}), the corresponding conformal transformation is given 
by
\begin{equation}
F(w) = \exp\{{ \textstyle\frac{1 + i}{\sqrt2}  w}\} .
\end{equation}
Using polar coordinates
so that $z = x + i y = r \exp({i \vartheta})$ we get\footnote{Note that in
RP, in equations (102--104), there is a misprint and the coordinate
$r$ must be substituted with its natural logarithm.}
\begin{equation}\label{coordiz2}
      X = \tfrac{1}{\sqrt2} (\vartheta + \ln r) ,\quad
      Y = \tfrac{1}{\sqrt2} (\vartheta - \ln r) .
\end{equation}
The $X, Y=constant$ curves are respectively given by the two families
\begin{equation}
r_1(\vartheta) = \exp (\sqrt2 X - \vartheta) ,\quad
           r_2(\vartheta) = \exp (\sqrt2 Y + \vartheta) .
\end{equation}
The range of $X,Y$ is
from $-\infty$ to $+\infty$ so that the punctured plane (with the origin
excluded) is covered with a one-to-one correspondence. 

%%%%%%%%%%%%%%%%%%%%%%%%%%%%%%%%%%%%%%%%%%%%%%%%%%%%%%%%%%%%%%%%%%%%%%%%%%%%%%

\subsection{A class of weakly integrable systems with $S(z) = i z^2$}

  From the treatment of section 2, it follows that the potential given by
\begin{equation}\label{iz2G}
        V(X,Y) = \frac{f(X) + g(Y)}{r^2} ,
\end{equation}
with the separating coordinates of (\ref{coordiz2}), is integrable at zero
energy for arbitrary functions $f$ and $g$.  The factor $ r^{-2} $ comes 
from
eq.(\ref{confact}) that, from the choice of eq.(\ref{iz2}), gives
\begin{equation}
|S(X,Y)| = {\rm e}^{\sqrt2 (X - Y)} = r^2 .
\end{equation}
Due to the presence of the polar angle as explicit argument in combination
with the logarithm of the radial coordinate, one must be careful in the
selection of the arbitrary functions to avoid unphysical multivaluedness in
the resulting potential function.  A simple choice giving a smooth glueing 
of
the sheets, resulting in a continuous and single-valued potential, is the
following
\begin{eqnarray}\label{iz2AB}
      f(X) &=& \tfrac{1}{2} (C - \sin {\sqrt2} X) ,\\ 
      g(Y) &=& \tfrac{1}{2} (C - \sin {\sqrt2} Y) ,
\end{eqnarray}
where $C$ is a real constant. Using (\ref{coordiz2}), the explicit form of 
the
potential in polar coordinates is
\begin{equation}\label{iz2pot}
       V(r, \vartheta) = \frac{C - \sin \vartheta \cos (\ln r)}{r^2} .
\end{equation}
When the parameter $C$ lies in the interval
\begin{equation}\label{Cvalues}
0 < C \le 1 ,
\end{equation}
the motion in the potential (\ref{iz2pot}) is bound for every value of the
energy below a positive threshold, since the equipotentials are closed 
regular
curves around the unique absolute minimum of the function $V(r, 
\vartheta)$,
\begin{equation}
E_{min} = {\rm min} \{V(r, \vartheta)\} < 0.
\end{equation}
At the threshold energy, $E_T > 0$, the equipotential is asymptotically 
open
and the motion is of parabolic type. We are interested in the bound 
motion, so
that we study the energy range
\begin{equation}
E_{min}  < E < E_T.
\end{equation}
At zero energy we can readily verify that the system is indeed integrable. 
Using
momenta expressed in polar coordinates according to
\begin{eqnarray}\label{polarmom}
 r p_r &=& x p_x + y p_y , \\
 p_{\vartheta} &=& x p_y - y p_x , 
\end{eqnarray}
the second invariant has the form
\begin{equation}\label{secondiz2}
I_0 (p_r, p_{\vartheta}, r, \vartheta) =
     r p_r p_{\vartheta} - \cos \vartheta \ \sin (\ln r) .
\end{equation}
It is straightforward to check that this function commutes with Hamiltonian
\begin{equation}\label{horig}
       {\cal H} =  \tfrac{1}{2} \left(p_r^2 + \tfrac{p_{\vartheta}^2}{r^2} \right)  + 
\frac{C - \sin \vartheta \cos (\ln r)}{r^2}  
\end{equation}
on the surface (\ref{zero}).

Moreover, a direct inspection allows to find the main periodic orbits. We 
have the
marginally stable orbit
\begin{equation}\label{r=1}
r \equiv 1
\end{equation}
and the two stable periodic orbits given by
\begin{equation}\label{peri1}
\vartheta = \pm \ln r .
\end{equation}

%%%%%%%%%%%%%%%%%%%%%%%%%%%%%%%%%%%%%%%%%%%%%%%%%%%%%%%%%%%%%%%%%%%%%%%%%%%%%%

\subsection{Dynamics outside the zero energy shell}

The interesting question is now: what is the nature of the motion when the
energy is different from zero? Since the system has been constructed as the
simplest generalization of separable systems, it could be still integrable 
possibly with a more complicated second invariant. A
systematic investigation of polynomial
invariants of higher degree has been unsuccessful. This is clearly not 
enough to
state non-integrability of system (\ref{horig}). However, we can take the
other way and try to see if we can devise a test of non-integrability. 
Actually, it
would be highly desirable to have an analytic test of non--integrability, 
since
it can escape a numerical approach, such as the computation of the 
Poincar\'e
surface, if stochastic zones are exceedingly small to be detected. 

An analytic proof can be
obtained by applying the Poincar\'e theorem
(Poincar\'e, 1892; Whittaker, 1937) on
non-existence of additional invariants in non-degenerate systems.  To this 
end
we follow a perturbative approach in which the value of the energy plays 
the
role of the perturbation parameter, with zero energy defining the 
unperturbed
system. We remark that this procedure can be applied to the regularized 
system
whereas it seems very difficult to attempt it in the original variables.

In the present instance, the null Hamiltonian (\ref{nullHam}) in the 
separating variables is
\begin{equation}\label{nullHamiz2}
{{\cal H}}_{\cal N} = \tfrac12 (p_X^2 + p_Y^2) -
              \tfrac12 \bigl(\sin {\sqrt2} X + \sin {\sqrt2} Y \bigr) + C -
               E \ {\rm e}^{\sqrt2 (X - Y)}.
\end{equation}
With the further change of variable
\begin{eqnarray}
\alpha(X) &=& {\sqrt2} X - \frac{\pi}{2} = \vartheta + \ln r -
\frac{\pi}{2},\\
\beta (Y) &=& {\sqrt2} Y - \frac{\pi}{2} = \vartheta - \ln r -
\frac{\pi}{2},\end{eqnarray}
we obtain the Hamiltonian
\begin{equation}\label{hp1}
{{\cal H}}_P = \tfrac12 (p_{\alpha}^2 + p_{\beta}^2) -
                         \cos \alpha - \cos \beta - 2 E {\rm e}^{\alpha -
\beta} = - 2 C.
\end{equation}
We can now express the system in the standard form $
   {{\cal H}}_P = {{\cal H}}_0 + \epsilon {{\cal H}}_1 $,
where an integrable ``zero" order Hamiltonian is perturbed by a 
non-integrable
first order term. In our case we have
\begin{eqnarray}\label{perturbedHam}
{{\cal H}}_0 &=& \tfrac12 (p_{\alpha}^2 + p_{\beta}^2) -
                          \cos \alpha - \cos \beta ,\\
{{\cal H}}_1 &=& {\rm e}^{\alpha - \beta} ,\\
\epsilon  &=& - 2 E .
\end{eqnarray}
It is then necessary to express this Hamiltonian as a function of the
action-angle
variables of the unperturbed system:
\begin{equation}
{\cal H}_P =  {\cal H}_0 (J_1, J_2) + \epsilon
                {\cal H}_1 (J_1, J_2, \theta_1, \theta_2) .
\end{equation}
The canonical variables satisfy the equations of motion
\begin{equation}
      J_k' = - \frac{\partial {{\cal H}}_P}{\partial \theta_k} ,\quad
      \theta_k' = \frac{\partial {{\cal H}}_P}{\partial J_k} ,\quad
      (k =1,2) \ ,
\end{equation}
where the prime denotes differentiation with respect to $\eta$, the new 
time
introduced in (\ref{newtime}).  For $E=0$ ($\epsilon = 0$), the unperturbed
motion is given by:
\begin{eqnarray}
      J_k'   = 0 & \rightarrow & J_k    = {\rm const},\\
      \theta_k' = \frac{\partial {{\cal H}}_0}{\partial J_k} = \omega_k =
      {\rm const} & \rightarrow & \theta_k = \omega_k \eta + {\rm const}.
\end{eqnarray}
The energy equations (\ref{energies}) represent in this case the two 
uncoupled
pendulums of the unperturbed Hamiltonian.  Now we have
\begin{equation}\label{hhC}
h_1 + h_2 = - 2 C.
\end{equation}
In view of (\ref{Cvalues}), we are considering only libration motions. It 
follows that the
action variables are (see, e.g., Boccaletti and Pucacco, 1996)
\begin{equation}\label{actionsjk}
J_k = \frac{8}{\pi} \bigl[ E(\kappa_k^2) + (\kappa_k^2 - 1)
                                 K(\kappa_k^2)\bigr],
\end{equation}
where
\begin{equation}\label{kkC}
\kappa_k^2 = \tfrac12 \bigl( 1 + h_k \bigr), \quad 0 \le \kappa_k^2 < 1
\end{equation}
and $K(\kappa_k^2)$ and $E(\kappa_k^2)$ are the complete elliptic 
integrals of
first and second kind.  ${{\cal H}}_0 $ is implicitly expressed in terms 
of the
actions by means of
\begin{equation}\label{hzp}
{\cal H}_0 (J_1, J_2) = 2 \bigl(\kappa_1^2 (J_1) + \kappa_2^2 (J_2) - 1 
\bigr).
\end{equation}

Also the main periodic orbits of the
unperturbed potential of Hamiltonian (\ref{hp1}) deserve to be mentioned. 
The
two periodic orbits
\begin{equation}\label{A.6}
\alpha =0,\quad
           \beta  =0,
\end{equation}
are elliptic stable and coincide with the two periodic orbits in
eq.(\ref{peri1}).  The two periodic orbits
\begin{equation}\label{A.7}
\alpha - \beta =0,
\end{equation}
and
\begin{equation}\label{A.8}
             \alpha + \beta =0,
\end{equation}
are of parabolic type (that is they are marginally stable with 
characteristic
exponents $\pm 1$).  In particular (\ref{A.7}) coincides with the orbit of
eq.(\ref{r=1}).  These orbits are ``resonant" since their frequencies
${\omega}_1, {\omega}_2$ are equal and are members of a family given by 
all the
possible phase shifts $\lambda$ with 
\begin{equation}
0 \le \lambda \le {{2 \pi} \over {{\omega}_1}}.  
\end{equation}
Orbit (\ref{A.7}) corresponds to
$\lambda=0$, whereas orbit (\ref{A.8}) corresponds to
$\lambda=\pi/{\omega}_1$. Parabolicity
is just related to the existence of resonant tori in the unperturbed 
system. A
perturbation ``breaks" these tori generating isolated stable and unstable
periodic orbits. In general, in non-degenerate systems we may expect the
existence of a generic resonance that in the following will be taken of the
form
\begin{equation}\label{reso}
{\omega}_1 = (p/q) {\omega}_2, 
\end{equation}
with $p$ and $q$ integers such that $p \le q$.  

%%%%%%%%%%%%%%%%%%%%%%%%%%%%%%%%%%%%%%%%%%%%%%%%%%%%%%%%%%%%%%%%%%%%%%%%%%%%%%

\subsection{Poincar\'e theorem on non-existence of additional
invariants }

Poincar\'e method to demonstrate non-integrability of a Hamiltonian 
systems is based 
on proving that, under some genericity conditions, no analytic second 
invariant
exist besides the Hamiltonian itself. We state the theorem as it is 
presented
in Poincar\'e (1892) and Whittaker (1937). Then we
show (theorem 5 below) that our system (\ref{nullHamiz2}) comply with the 
requirements of
Poincar\'e theorem, proving therefore its non-integrability and, as a 
consequence,
non-integrability of original system (\ref{horig}). Moreover, by 
exploiting lemma 4
used in the proof of theorem 5, we will show in the following section the 
existence of isolated
periodic orbits and verify that their stable (or unstable) nature varies 
with respect to $E$.

We start with a definition. Assume that, for $J_1 = {\hat J}_1$ and $J_2 = 
{\hat J}_2$,
frequencies ${\hat \omega}_1$ and ${\hat \omega}_2$ are commensurate 
according
to relation (\ref{reso}). 
$J_1, J_2$ satisfying this condition are said to belong to the {\it 
Poincar\'e
set}. For a two dimensional system, ``Poincar\'e non-existence theorem" 
can be stated as
follows:

\vskip.3cm

\noindent
{\bf Theorem 2:} Let the Hamiltonian
\begin{equation}
{\cal H}_P = {\cal H}_0 (J_1, J_2) + \epsilon
             {\cal H}_1 (J_1, J_2, \theta_1, \theta_2) 
\end{equation}
satisfy the following hypotheses:

1. 
${\cal H}_0 $ is non-degenerate, that is:
\begin{equation}\label{condition1}
{\rm det} \biggl(\frac{\partial^2 {\cal H}_0}{\partial {J}_k^2}
\biggr) \ne 0 ;
\end{equation}

2. 
${\cal H}_1 $ is generic, that is, defining the Fourier expansion
\begin{equation}
{\cal H}_1 (J_1, J_2, \theta_1, \theta_2) =
\sum_{m,n=-\infty}^{+\infty} h_{mn} (J_1, J_2) {\rm e}^{i(m \theta_1 + n
\theta_2)},
\end{equation}
no coefficient $h_{mn} (J_1, J_2)$ is zero in the Poincar\'e
set;

\noindent
then, there is no analytic second invariant of the form
\begin{equation}\label{asi}
I = \sum_{s=0}^{\infty} \epsilon^s I_{s} (J_1, J_2, \theta_1 , \theta_2),
\end{equation}
 independent of ${\cal H}$.

\vskip.3cm

{\it Proof:} We refer to Poincar\'e (1892, sects. 81--83) and Whittaker 
(1937,
sect. 165). For a more recent proof, see Giorgilli (2002, sect.3.1). Q.E.D.

\vskip.3cm

To show the non existence of analytic second invariants of the form 
(\ref{asi}) is
equivalent to exclude Liouville integrability. However, it must be 
remarked that the above
result apply in the case of a regular function which is uniformly 
continuous in the
phase-space variables and in a small range of the perturbation parameter, but the possibility
cannot be excluded that for small fixed values of $\epsilon$ does a 
non-uniform invariant
indeed exist (see, e.g., Kozlov, 1991, sect. 4.1).

We now proceed to verify that Hamiltonian (\ref{hp1}) satisfy the
conditions required to apply theorem 2. Let us check the non-degeneracy
condition first.
 
\vskip.3cm

\noindent
{\bf Lemma 3:} The Hamiltonian (\ref{hzp}) is non-degenerate.

\vskip.3cm

{\it Proof:} The frequencies of the unperturbed motion can be calculated as
follows (see, Boccaletti and Pucacco (1996), sect. 1.16):
\begin{equation}\label{A.1}
\omega_k = \frac{\partial {{\cal H}}_0}{\partial J_k} =
            \frac{\partial}{\partial J_k} \bigl(2 \kappa_k^2 \bigr) =
4 \kappa_k \biggl(\frac{\partial J_k}{\partial\kappa_k}\biggr)^{-1} =
\frac{\pi}{2} \frac{1}{K(\kappa_k^2)},
\end{equation}
where we have used
\begin{equation}\label{A.2}
\frac{\partial J_k}{\partial\kappa_k} =
\frac{8}{\pi} \kappa_k K(\kappa_k^2),
\end{equation}
which comes from eq.(\ref{actionsjk}) and standard properties of the 
elliptic
integrals. With an analogous procedure we can compute the second derivative
\begin{equation}\label{A.3}
\frac{\partial^2 {{\cal H}}_0}{\partial J_k^2} =
\frac{\partial \kappa_k}{\partial J_k}
\frac{\partial^2 {{\cal H}}_0}{\partial \kappa_k \partial J_k} =
4 \biggl(\frac{\partial J_k}{\partial \kappa_k}\biggr)^{-1}
\frac{\partial \kappa_k}{\partial J_k} =
\frac{16}{\pi^2} \frac{1}{\kappa_k^2 K^2(\kappa_k^2)}.
\end{equation}
We recall that $K(\kappa^2)$ is a positive uniformly increasing function 
in the interval $0 \le
\kappa^2 < 1$, such that
\begin{equation}\label{KK}
K(0) = {\pi \over 2}, \quad K(1) \rightarrow \infty.
\end{equation}
Since
\begin{equation}\label{A.4}
\frac{\partial^2 {{\cal H}}_0}{\partial J_1 \partial J_2} = 0,
\end{equation}
we have that
\begin{equation}
{\rm det} \biggl(\frac{\partial^2 {{\cal H}}_0}{\partial {\hat J}_k^2}
    \biggr)= \frac{\partial^2 {{\cal H}}_0}{\partial {\hat J}_1^2}
             \frac{\partial^2 {{\cal H}}_0}{\partial {\hat J}_2^2} \ne 0,
\end{equation}
so that eq.(\ref{condition1}) is satisfied and condition 1. in the 
statement of
Theorem 2 is verified. Q.E.D.

\vskip.3cm

We come now to the genericity condition, directly proceeding to perform a 
Fourier
series expansion of the perturbation appearing in (\ref{hp1}):
\begin{equation}\label{fourierexp}
{\cal H}_1 = {\rm e}^{\alpha(\eta) - \beta(\eta)} =
\sum_{m,n=-\infty}^{+\infty} h_{mn} (J_1, J_2) {\rm e}^{i(m \theta_1 + n 
\theta_2)},
\end{equation}
where we have to use the solution of the unperturbed problem. The motion 
in the 
potential of Hamiltonian (\ref{hp1}) with $E=0$ is
given by 
\begin{eqnarray}
\label{alpha}\alpha(\eta) &=& 2
\arcsin \bigl[ \kappa_1 {\rm sn} (\eta,
\kappa_1)\bigr],\\
\label{beta}\beta(\eta) &=& 2
\arcsin \bigl[ \kappa_2 {\rm sn} (\eta,
\kappa_2)\bigr],
\end{eqnarray}
where ${\rm sn}(u)$ is the Jacobi sine-amplitude function. In this way, 
since $\kappa$
depends on $J$, the actions of the unperturbed problem enter in the 
expansion. On the same
footing, the angles appearing in the Fourier expansion are just the 
canonical angle variables of
the unperturbed problem, so that the harmonics
$m \omega_1, n \omega_2$, with integer $m,n$, are multiples of the 
unperturbed
frequencies. 

We preliminarly assess the structure of the resonating set of frequencies 
that determine the
Poincar\'e set. Using expressions (\ref{A.1}) for the frequencies, we have 
that the
resonance condition (\ref{reso}) can be written as
\begin{equation}\label{resoKK}
{{q} \over {K(\kappa_1^2)}} - {{p} \over {K(\kappa_2^2)}}  = 0,
\end{equation}
where, in view of (\ref{hhC}) and (\ref{kkC}),
\begin{equation}\label{CCKK}
\kappa_2^2 = 1 - C - \kappa_1^2.
\end{equation}
Choosing a value of the parameter $C$ in the interval (\ref{Cvalues}) 
fixes the corresponding
range of
$\kappa_1^2$ as
\begin{equation}\label{K1range}
0 \le \kappa_1^2 \le 1 - C.
\end{equation}
Therefore, we can define a function
\begin{equation}\label{resoRR}
R(\kappa_1^2) = {{q} \over {p}} K(1 - C - \kappa_1^2) - K(\kappa_1^2),
\end{equation}
which is continuous in the range (\ref{K1range}) and, recalling 
(\ref{KK}), such that
\begin{equation}\label{limitRR}
R(0) = {{q} \over {p}} K(1 - C) - {\pi \over 2}, \quad 
R(1-C) = {{ \pi q} \over {2p}} - K(1 - C).
\end{equation}
Without loss of generality we have assumed $p \le q$. Recalling again 
(\ref{KK}), we have that
$$
K(1 - C) > {\pi \over 2}
$$
which implies $R(0) > 0$. Moreover, if the inequality
\begin{equation}\label{pqine}
{{q} \over {p}} < {{2} \over {\pi}} K(1 - C)
\end{equation}
is satisfied, we see that $R(1-C) < 0$. Then we have that the function 
$R(\kappa_1^2)$ defined 
in the interval (\ref{K1range}), by continuity, must necessarily vanish at 
a point of this
interval. Observing that condition ({\ref{resoKK}) can also be written in 
the
form
\begin{equation}\label{resoKK2}
{{{{q} \over {p}} K(1 - C - \kappa_1^2) - K(\kappa_1^2)} \over 
{K(1 - C - \kappa_1^2) K(\kappa_1^2)}}  = 0,
\end{equation} 
we see that the point where $R(\kappa_1^2)$ vanishes is also the point 
where the resonance
condition is satisfied. 

Actually, it is
straightforward to check that inequality ({\ref{pqine}) is not a 
limitation on the set of
possible resonances, since dynamics automatically comply with it. In fact, 
recalling again
(\ref{A.1}) and {\ref{KK}), we see that both frequencies of the 
unperturbed libration motions
must satisfy the constraints
\begin{equation}\label{freqlimit}
{{\pi} \over {2 K(1 - C) }} = \omega (1-C) < \omega < \omega(0) = 1.
\end{equation}
Therefore, the  ratio of two frequencies is such that
\begin{equation}\label{freqratio}
{{\omega_2} \over {\omega_1}} < {{\omega (0)} \over {\omega(1-C)}} = {{2} 
\over {\pi}} K(1 - C),
\quad \forall \ C \in (0,1].
\end{equation}
As a consequence we see that, for every pair of frequencies such that the 
resonance condition
({\ref{reso}) is verified, inequality (\ref{pqine}) is always satisfied. 
We conclude that,
{\it for every value of the parameter $C$} in the range (\ref{Cvalues}), 
there is a whole dense
set of resonances in the space of unperturbed frequencies. Each value of 
the root of the
function $R(\kappa_1^2)$, say ${\hat \kappa}_1^2 (C,q,p)$ with the 
corresponding value of
${\hat \kappa}_2^2$ given by (\ref{CCKK}), determine, via 
(\ref{actionsjk}), the
corresponding values of ${\hat J}_1$ and ${\hat J}_2$ of the Poincar\'e
set.

At this point, we have the following:

\vskip.3cm

\noindent
{\bf Lemma 4:} 
The Fourier expansion coefficients of perturbation (\ref{fourierexp}) are
non-vanishing on the resonance manifold specified by relation (\ref{reso}).

\vskip.3cm

{\it Proof:} The Fourier coefficients are given by 
\begin{equation}\label{fourierhmn}
h_{mn} = A_m B_n, 
\end{equation}
where
\begin{equation}\label{fourierca}
A_m = \frac{1}{T_1} \int_0^{T_1} {\rm e}^{\alpha(\eta)} 
                {\rm e}^{-i m \omega_1 \eta} d \eta,
\end{equation}
\begin{equation}\label{fouriercb}
B_n = \frac{1}{T_2} \int_0^{T_2} {\rm e}^{-\beta(\eta)} 
                {\rm e}^{-i n \omega_2 \eta} d \eta,
\end{equation}
and, in a natural way,
$$
T_k = \frac{2\pi}{\omega_k}, \ \ \ k = 1,2.
$$
On the resonance manifold we have
\begin{equation} {m \over n} = -{q \over p},
\end{equation}
so that the simplest choice we can make is 
\begin{equation}m=q, \ \ \ n=-p.
\end{equation}

Let us separate real and
imaginary parts in the Fourier coefficients of a generic resonance.
The coefficient with $m=q$ is:
\begin{equation}\label{fraia}
A_q = \frac{1}{T_1} \int_0^{T_1} {\rm e}^{\alpha(\eta)}
     (\cos q {\hat \omega}_1 \eta - i \sin q {\hat \omega}_1 \eta)
                d \eta = C_q - i S_q.
\end{equation}
Using the resonance condition
$$
n {\hat \omega}_2 = - p {\hat \omega}_2 = - q {\hat \omega}_1,
$$
the coefficient with $n=-p$ is:
\begin{equation}\label{frbib}
B_{-p} = \frac{q}{pT_1} \int_0^{\frac{p}{q}T_1} {\rm e}^{-\beta(\eta)}
     (\cos q {\hat \omega}_1 \eta + i \sin q {\hat \omega}_1 \eta)
                d \eta = C_p + i S_p.
\end{equation}
The Fourier coefficients on the resonance manifold are therefore
\begin{equation}\label{frrm}
h_{q,-p} = (C_q C_p +S_q S_p) + i (C_q S_p - C_p S_q).
\end{equation}
We now proceed to evaluate above integrals, starting with the elimination 
of all cases in which
some of them identically vanish. First of all, observing solutions
(\ref{alpha}--\ref{beta}), we realize that the functions $\alpha(\eta)$ 
and $\beta(\eta)$ are
even over each {\it half} oscillation period and odd over the 
corresponding {\it whole} period.
As a consequence, the exponentials appearing in the perturbation 
(\ref{fourierexp}) are both
even functions over the corresponding half period. On the same interval, 
$\cos q {\hat
\omega}_1 \eta$ (with $q$ odd) and $\sin q {\hat \omega}_1 \eta$ (with $q$ 
even) are
odd functions. Referring to the integrals in
\eqref{fraia}, by parity arguments, we can therefore state that:
\begin{eqnarray}\label{cqsq}
         C_q &=& 0 \;\; {\rm for} \; q \; {\rm odd}, \\
         S_q &=& 0 \;\; {\rm for} \; q \; {\rm even}.
\end{eqnarray}
The remaining integrals, namely $S_q$ with $q$ odd and $C_q$ with $q$ 
even, are surely non
vanishing. For, take for instance $S_q$ when $q$ is odd. We have
\begin{eqnarray}
T_1 S_q &=& 
\int_0^{\frac{1}{2}T_1}     {\rm e}^{\alpha(\eta)} \sin q {\hat \omega}_1 
\eta d \eta +
\int_{\frac{1}{2}T_1}^{T_1} {\rm e}^{\alpha(\eta)} \sin q {\hat \omega}_1 
\eta d \eta \\
 &=& 
\int_0^{\frac{1}{2}T_1}     {\rm e}^{ \alpha(\eta)} \sin q {\hat \omega}_1 
\eta d \eta -
\int_0^{\frac{1}{2}T_1}     {\rm e}^{-\alpha(\eta)} \sin q {\hat \omega}_1 
\eta d \eta,
\end{eqnarray}
where, in the second step, we performed a translation $\eta \rightarrow 
\eta - T_1/2$ and
used the fact that
\begin{eqnarray}
\alpha(\eta - T_1/2) &=& - \alpha(\eta) \\
\sin q {\hat \omega}_1 (\eta - T_1/2) &=& - \sin q {\hat \omega}_1 \eta.
\end{eqnarray}
The statement follows because 
$${\rm e}^{ \alpha( \eta)} \ne {\rm e}^{ -\alpha(\eta)}, \ \forall \ \eta 
\ne 0.$$
The same
holds for $C_q$ when $q$ is even.

The integrals appearing in \eqref{frbib} are
characterized by the integration interval which depends on the ratio 
$p/q$. However, the
function ${\rm exp}(-\beta)$, which has the same parity property of ${\rm 
exp}(-\alpha)$, is
``sampled" by trigonometric functions alternatively even and odd on each 
half interval. Using
the same arguments as above we get
\begin{eqnarray}\label{cpsp}
         C_p &=& 0, \;\; S_p \ne 0, \;\; {\rm for} \; p \; {\rm odd}, \\
         S_p &=& 0, \;\; C_p \ne 0, \;\; {\rm for} \; p \; {\rm even}.
\end{eqnarray}

Summarizing these results and comparing with expression (\ref{frrm}), we 
get finally that the
following coefficients 
\begin{eqnarray}
h_{q,-p} &=&    C_q C_p \;\; {\rm for} \; q,p \; {\rm even},\\
h_{q,-p} &=&  i C_q S_p \;\; {\rm for} \; q \;   {\rm even} \; {\rm and} 
\; p \; {\rm odd},\\
h_{q,-p} &=& -i S_q C_p \;\; {\rm for} \; q \;   {\rm odd}  \; {\rm and} 
\; p \; {\rm even},\\
h_{q,-p} &=&    S_q S_p \;\; {\rm for} \; q,p \; {\rm odd},
\end{eqnarray}
are non vanishing. This proves the lemma and shows that condition 2. in 
the statement of theorem
2 is indeed satisfied. Q.E.D.

\vskip.3cm

We remark that, according to the analysis already carried out by 
Poincar\'e (1892, sect.83),
condition 2. can be somewhat relaxed. In particular, introducing the 
concept of
``classes" in the resonance manifold, the first combination with both $q$ 
and $p$ even can be
reduced to one of the others. However, this fact does not substantially 
modify the 
conclusions above.  

With this body of results we can now enunciate our non-integrability 
statement applied to the
original class of systems of section 3.1.

\vskip.3cm

\noindent
{\bf Theorem 5:} The dynamical system generated by Hamiltonian,
\begin{equation}\label{hamilto}
       {\cal H} =  \tfrac{1}{2} \left(p_r^2 + \tfrac{p_{\vartheta}^2}{r^2} 
\right) + \frac{C - \sin \vartheta \cos (\ln r)}{r^2} = E,  
\end{equation}
integrable at energy $E=0$, does not admit a conserved function
\begin{equation}
I(p_r,p_\vartheta,r,\vartheta; C),
\end{equation}
defined in the open set 
$-\epsilon_1 < E < 0$ and $0 < E < \epsilon_2$, with small enough 
$\epsilon_{1,2}$, for every value of $C$ in the interval  $(0,1]$.

\vskip.3cm

{\it Proof:} From the discussion on regularization and dynamics at zero 
energy of sects. 2.3
and 2.4, it has been shown that the dynamics generated by Hamiltonian 
(\ref{hamilto}) is
equivalent, a part a time reparametrization, to those generated by 
Hamiltonian
(\ref{hp1}). In particular, this applies to the conservation properties of 
phase-space
functions: non-existence of a function $I(p_\alpha,p_\beta,\alpha,\beta; E) $
which is an invariant, for any $E$, of Hamiltonian (\ref{hp1}),
implies non-existence of a function $I(p_r,p_\vartheta,r,\vartheta; C)$
which is an invariant, for any $C$, of Hamiltonian
(\ref{hamilto}). Hamiltonian (\ref{hp1}) satisfies the conditions for the 
applicability of
theorem 2. For, by lemma 3 we have that its unperturbed part is non 
degenerate and from lemma 4
we have that the perturbing function is generic in the Poincar\'e set 
which is dense for every
$0 < C \le 1$. The statement follows. Q.E.D.

\vskip.3cm

We recall (see sect. 3.1) that Hamiltonian (\ref{hamilto}) supports bound 
motion in the energy
range 
\begin{equation}
E_{min}  < E < E_T
\end{equation}
with $ E_{min} < 0 $ and $ E_T > 0 $ and is separable at $E=0$. Theorem 5 
proves
non-integrability of these motions for small (positive or negative) values 
of the energy.
However, from the remarks above, it is not possible to exclude the 
existence of isolated
integrable case with non uniform invariants corresponding to some other 
fixed values of the energy.

%%%%%%%%%%%%%%%%%%%%%%%%%%%%%%%%%%%%%%%%%%%%%%%%%%%%%%%%%%%%%%%%%%%%%%%%%%%%%%%%

\subsection{Existence of isolated periodic orbits}

Non-integrability of a perturbed Hamiltonian system manifests itself with 
the birth of
alternately stable and unstable periodic orbits. Actually, their existence 
is an
``obstruction" to the integrability and can be used as an argument to 
prove non-integrability
when the conditions required by theorem 2 do not apply or are difficult to 
be checked. In our
case, we can exploit the results already obtained to get informations 
about the nature of the
periodic orbits ensuing from the perturbation. We follow
Arnold, Kozlov and Neishtadt (1988, sect. 6.1) to show the existence of 
isolated
periodic orbits and to determine their stability properties. For a recent 
application of this
technique in celestial mechanics we refer to Diacu and Santoprete (2001). 

We state, referring for the proof again to Poincar\'e (1892), the 
following

\vskip.3cm

\noindent
{\bf Theorem 6:} Suppose that the following conditions are satisfied:

\vskip.3cm

1.
${{\cal H}}_0 $ is non-degenerate in ${\hat J}_k, \, k=1,2$, that is:
\begin{equation}
{\rm det} \biggl(\frac{\partial^2 {{\cal H}}_0}{\partial {\hat J}_k^2}
\biggr) \ne 0 ;
\end{equation}

2.
for some $\lambda = \lambda^{\star}$, the average perturbation,
defined as
\begin{equation}\label{ave}
{\overline {{\cal H}}_1} ({\hat J}_1, {\hat J}_2, \lambda) =
\frac{1}{T_{\eta}} \int_0^{T_{\eta}}
   {{\cal H}}_1 ({\hat J}_1, {\hat J}_2, {\hat \omega}_1 \eta + \lambda,
{\hat \omega}_2 \eta) d \eta ,
\end{equation}
is such that
\begin{equation}\label{condition2}
\frac{\partial {\overline {{\cal H}}_1}}{\partial\lambda} = 0 ,\quad
\frac{\partial^2 {\overline {{\cal H}}_1}}{\partial\lambda^2} \ne 0 ;
\end{equation}

and the perturbing
function 
\begin{equation}
{{\cal H}}_1 = {{\cal H}}_1 ({\hat J}_1, {\hat J}_2, {\hat \omega}_1 \eta 
+ \lambda,
{\hat \omega}_2 \eta) ,
\end{equation}
where $\lambda$ is the phase shift of the resonating motions, is periodic 
in
$\eta$ with period $T_{\eta}$. 

Then, for small
$\epsilon$, the perturbed system has a periodic solution with
period $T_{\eta}(\epsilon)$ such that $T_{\eta}(0)=T_{\eta}$ and with
characteristic exponents $\pm \alpha$ which can be expressed as power 
series
in the form
\begin{equation}\label{car}
\alpha = \alpha_1 \sqrt{\epsilon} +
           \alpha_2 \epsilon + \alpha_3 \epsilon \sqrt{\epsilon} + dots ,
\end{equation}
where the coefficients are given by the relations
\begin{equation}\label{leading}
{\hat \omega}_1^2 \alpha_1^2 =
\frac{\partial^2 {\overline {{\cal H}}_1} (\lambda^{\star})}{\partial 
\lambda^2}
\biggl(
{\hat \omega}_1^2 \frac{\partial^2 {{\cal H}}_0}{\partial {\hat J}_1^2} - 2
{\hat \omega}_1 {\hat \omega}_2
\frac{\partial^2 {{\cal H}}_0}{\partial {\hat J}_1 \partial {\hat J}_2} +
{\hat \omega}_2^2 \frac{\partial^2 {{\cal H}}_0}{\partial {\hat J}_2^2}
   \biggr) .
\end{equation}

\vskip.3cm

We can actually prove that system (\ref{perturbedHam}) meets the 
requirement of theorem 6.
For, condition 1. is a straightforward application of lemma 3 to the 
Poincar\'e set.
Condition 2. derives applying the results of lemma 4. Taking into account 
the phase shift, the
Fourier expansion of the perturbation can now be written in the form
\begin{equation}\label{fexp}
{\cal H}_1 (J_1, J_2, \eta, \lambda) = 
\sum_{m,n} {\tilde A}_m (\lambda) B_n {\rm e}^{i(m \omega_1 + n 
\omega_2)\eta},
\end{equation}
where
\begin{equation}\label{fca}
{\tilde A}_m (\lambda) = \frac{1}{T_1} {\rm e}^{i m \omega_1 \lambda} 
                \int_0^{T_1} {\rm e}^{[\alpha(\tau) - i m \omega_1 \tau]} 
                d \tau = {\rm e}^{i m \omega_1 \lambda} A_m
\end{equation}
and
$$
\tau = \eta + \lambda,
$$
In order to verify condition 2. we have to perform the average \eqref{ave} 
of ${\cal H}_1$.
Using the expansion \eqref{fexp}, we realize that the only $\lambda$ 
dependent terms surviving
the averaging process, are just those corresponding to resonances of the 
form \eqref{reso}, that
is 
\begin{equation}\label{fexpsurv}
{\cal H}_1 ({\hat J}_1, {\hat J}_2, \eta, \lambda) = 
\sum_{m,n} {\tilde A}_m (\lambda) B_n {\rm e}^{i(m + n \frac{q}{p}) {\hat 
\omega}_1 \eta}.
\end{equation}
Therefore, after averaging, we get terms of the form
\begin{equation}\label{A.9}
{\overline {{\cal H}}_1} ({\hat J}_1, {\hat J}_2, \lambda) = 2 {\rm Re}
\left\{{\rm e}^{i q {\hat \omega}_1 \lambda} A_q B_{-p}\right\}.
\end{equation}
With the same notation used in the proof of lemma 4, first and second 
derivative of ${\overline
{\cal H}_1} ({\hat J}_1, {\hat J}_2, \lambda) $ with respect to $\lambda$ 
are:
\begin{eqnarray}\label{deriv}
\frac{\partial {\overline {\cal H}_1}}{\partial \lambda} &=& 2 {\rm Re}
\left\{
i q {\hat \omega}_1 \biggl[ {\rm e}^{i q {\hat \omega}_1 \lambda}
(C_q + i S_q)(C_p + i S_p) \biggr] \right\},\\
\frac{\partial^2 {\overline {\cal H}_1}}{\partial \lambda^2}  &=& 2 {\rm Re}
\left\{
- q^2 {\hat \omega}_1^2 \biggl[ {\rm e}^{i q {\hat \omega}_1 \lambda}
(C_q + i S_q)(C_p + i S_p) \biggr] \right\}.
\end{eqnarray}
Therefore, condition 2. is satisfied if a $\lambda = \lambda^{\star}$ 
exists such that the
imaginary part of the term in the square bracket vanishes with a non 
vanishing real part. The
imaginary part is
\begin{equation}\label{ip}
\cos q {\hat \omega}_1 \lambda (C_q S_p + C_p S_q) + 
\sin q {\hat \omega}_1 \lambda (C_q C_p - S_q S_p).
\end{equation}
This expression vanishes if $\lambda$ takes the value
\begin{equation}\label{lambdastar}
\lambda^{\star} = \frac{1}{q {\hat \omega}_1} 
\arctan \frac{C_q S_p + C_p S_q}{S_q S_p - C_q C_p}.
\end{equation}
Correspondingly, we have:
\begin{equation}\label{secder}
\frac{\partial^2 {\overline {\cal H}_1}}{\partial \lambda^2} 
\bigg\vert_{\lambda = \lambda^{\star}} =
- 2 q^2 {\hat \omega}_1^2 \cos q {\hat \omega}_1 \lambda^{\star} 
[C_q (C_p + S_p) + S_q (C_p - S_p)].
\end{equation}

Using the results summarized in (\ref{cqsq}) and (\ref{cpsp}), if $n$ 
denotes an arbitrary
integer, we get that the critical shifts are either
\begin{equation}\label{lambd}
\lambda^{\star} = \frac{n \pi}{q {\hat \omega}_1}
\end{equation}
in case of an odd-odd or even-even $q,p$ combination, or
\begin{equation}
\lambda^{\star} = \frac{n \pi}{2 q {\hat \omega}_1} 
\end{equation}
in case of an even-odd or odd-even $q,p$ combination.

As a simple application, let
us take the simplest case, that is 
$$
p = q = 1, \ \ \ {\hat \omega}_1 = {\hat \omega}_2,
$$
so that the two unperturbed motions are identical but for the phase shift. 
From
\eqref{lambd} we then have
$$
\lambda^{\star} = \frac{n \pi}{{\hat \omega}_1}.
$$
In particular, 
\begin{equation}\label{A.12}
\lambda^{\star} = 0 \rightarrow
\frac{\partial^2 {\overline {\cal H}_1}}{\partial \lambda^2} 
= {\hat \omega}_1^2 S_{q=1}^2 > 0,
\end{equation}
so that, for the zero phase shift corresponding to the unperturbed 
periodic orbit \eqref{A.7},
the leading order term in the expansion (\ref{car}) of the characteristic 
exponent is a real
number. This can be seen using eq.(\ref{leading}) with the result 
\eqref{A.12} above and with
the round bracket that is positive in virtue of (\ref{A.3}) and 
(\ref{A.4}). Therefore, the
characteristic exponents are real if
$E<0$ and pure imaginary if
$E>0$.  The periodic orbit is then unstable in the former case and stable 
in the second. This
change in the nature of a periodic orbit on varying the energy is another 
typical
feature of non--integrable dynamics.

%%%%%%%%%%%%%%%%%%%%%%%%%%%%%%%%%%%%%%%%%%%%%%%%%%%%%%%%%%%%%%%%%%%%%%%%%%%%%%%%

\section{Conclusions}

The main points which we have focused on in the present paper can be
summarized as follows:

--- There exist wide classes of natural conservative mechanical systems 
with
integrable (actually separable) dynamics at energy equal to zero.

--- In the general case, the dynamics of the system is non-integrable at
energies different from zero.

--- The technique exploited here to construct these systems, which is 
based on
a generic conformal transformation, offers the possibility of 
characterizing
many features of the non-integrable dynamics.

In particular, we have seen how the structure of the transformed 
Hamiltonian
naturally allows a perturbative approach to investigate the departure from 
the
integrable regime.  Standard tools of analytical mechanics, like the
Poincar\'e methods, can be applied more easily than in the usual setting, 
to
get analytical proofs of non-integrability.

A natural continuation of this work is to explore the character of other
systems constructed by means of other conformal transformations or even of
other potentials in the classes introduced above.  In particular, we have
limited the analysis to systems admitting bound motion, but in the class 
of systems studied
above, it is easy to construct potentials allowing the coexistence of
limited and unlimited motions.  In such systems, dynamics are 
characterized by a
transition to chaotic scattering with possible applications in celestial 
mechanics, quantum
mechanics and general relativity.

\section{Acknowledgements}
We acknowledge the remarks of the referee, Dr. A. Albouy, that played a 
substantial role
in producing a clearer presentation of the results.

\end{document}